\documentstyle[osa,manuscript,psfig]{revtex}

\begin{document}                                                          
\titlepage

\title{Lambda polarization and single-spin left-right asymmetry 
in diffractive hadron-hadron collisions}
\author {Liang Zuo-tang$^1$ and C. Boros$^2$\\
 $^1$ Department of Physics, Shandong University, 
  Jinan, Shandong 250100, China\\
$^2$ Special Research Centre for the Subatomic Structure of Matter,
University of Adelaide, Adelaide, Australia 5005}
\maketitle    
                                                       
\begin{abstract}

We discuss Lambda polarization and 
single-spin left-right asymmetry  
in diffractive hadron-hadron scattering 
at high energies. 
We show that the physical picture proposed in a recent Letter 
is consistent with the experimental observation that 
$\Lambda$ polarization in the diffractive 
process, $pp\rightarrow \Lambda K^+p$, is much higher 
than that in the inclusive
reaction, $pp\to \Lambda X$. 
We make predictions for the  left-right asymmetry, $A_N$,  
and for the  spin transfer, $D_{NN}^\Lambda$, in    
the single-spin process $p(\uparrow)p\to \Lambda K^+p$ 
and suggest further experimental tests of the proposed picture.  

\end{abstract}

\newpage
              
It has been known \cite{Hel96} for a long time that 
hyperons produced in hadron-hadron 
or hadron-nucleus collisions 
are polarized transversely 
to the production plane, although 
neither the projectile nor the target 
is polarized before the collision.
Significant hyperon polarizations 
(up to 40\%) in inclusive production processes 
have been observed\cite{Hel96} in the fragmentation regions 
for moderately large transverse momenta. 
The effect has been confirmed 
in past years by a large number of similar
experiments at different energies 
and/or using different projectiles and/or targets and for 
the production of different  hyperons.  
However, the  physical origin of these striking polarizations   
is still a puzzle.   
The basic difficulty is that  
helicity of the almost massless quarks is conserved in  
perturbative QCD (pQCD) 
in leading twist and at leading order \cite{Kane78}, 
but the existence of large hyperon polarization 
requires significant helicity flip at the hadron level.   
Different attempts have been made \cite{HeliF} 
to overcome this difficulty in the context of pQCD, 
other theoretical  models have also been 
proposed [\ref{And79}-\ref{LB97}].   
It is clear that understanding of this striking 
spin effect would also shed light on 
the spin structure of hadron. 

While most of the experimental data on hyperon polarization are in  
inclusive processes, the  CERN R608 Collaboration 
has carried out an experiment \cite{R608} in  
the diffractive dissociation process 
$pp\to \Lambda K^+p$. 
One of the obvious advantages of this 
exclusive experiment is that, here, 
one concentrates on a much simpler final state 
than in inclusive processes. 
In this way, one hopes to gain deeper 
insight into the mechanisms of Lambda polarization.  
The results\cite{R608} of this experiment show that  
$\Lambda$ produced in this process 
is also transversely polarized,  
and that the polarization has the same sign (negative) 
as that observed\cite{Hel96} 
in the inclusive process $pp\to \Lambda X$ 
but the magnitude is very large
$(62\% \pm 4\%)$ --- much larger than those observed 
in the inclusive process.  
Since the diffractive process $pp\to \Lambda K^+p$ 
is the simplest channel for 
the inclusive process $pp\to \Lambda X$,   
the observation that $\Lambda$ polarization $P_\Lambda $ 
in this channel has a much larger value than  
that in $pp\to \Lambda X$, which is the 
average over all the different channels,  
suggests that this process plays indeed  
a special role.

In a recent Letter \cite{LB97}, we argued that 
there is a close relation between the above mentioned 
hyperon polarization ($P_H$) observed\cite{Hel96} 
in inclusive production processes in 
unpolarized hadron-hadron collisions 
and the left-right asymmetry ($A_N$) observed\cite{Bra98} 
in inclusive hadron-hadron collisions 
using transversely polarized projectile or target. 
Theoretical arguments and experimental observations 
have been presented which strongly suggest 
that the two phenomena have the same origin 
and should be considered together. 
If this is indeed the case, 
it offers a new starting point to understand  
the origin of $P_H$. 

In this note, 
we apply the picture to $\Lambda$ production in 
diffractive hadron-hadron collisions to make further 
tests of the picture by 
comparing the obtained results with the available data. 
We show in particular that the much larger value of $P_\Lambda$ 
in $pp\to \Lambda K^+p$ should be considered as 
a further strong evidence for the picture.  
We make suggestions for future experiments. 

We now start by recollecting the key points 
of the picture proposed in [\ref{LB97}].
The basic idea of the picture is 
that there is a close relation between 
$P_H$ in unpolarized hadron-hadron collisions  
and $A_N$ in single-spin hadron-hadron collisions. 
Hence, if we extract the essential points 
encoded in the $A_N$ data, 
we should be able to understand $P_H$ based 
on these points.    
The following two points, (a) and (b), have therefore 
been derived from  the existing  $A_N$ data\cite{Bra98}, 
and used as inputs to study $P_H$.

(a) {\it If a hadron 
is produced by an upwards polarized valence quark of the 
transversely polarized projectile, 
it has a large probability to have a transverse momentum
pointing to the ``left-hand side'' looking down stream. 
This production mechanism  
gives rise  to  positive  left-right asymmetry.} 
\cite{Bra98,E704a} 
Here,  ``left-hand side'' is defined  by the requirement 
that the scalar product, 
${\cal M}=\vec{S}\cdot (\vec{p}_B\times \vec{p})$, 
be positive, where 
$\vec{S}$  is  the polarization vector of the
transversely polarized  beam, $\vec{p}_B$ and  
$\vec{p}$ are the momenta of the beam and the 
produced hadron, respectively. 
Positive $A_N$ measures the excess of hadrons produced 
to the left-hand side over those produced to the right-hand side.  
The above point has been derived directly from 
the data\cite{Bra98,E704a} on $A_N$ 
in meson production. We denote the difference 
of the probabilities for ${\cal M}$ to be positive 
and  to be negative by $C$. $C$ 
should lie in the region $0<C<1$.    

(b) {\it If a hadron
is produced by two valence quarks 
(valence diquark) of the projectile, 
the remaining valence quark produces 
an associated hadron. 
The left-right asymmetry 
due to this production mechanism  
is opposite to that of the associated hadron. }
This is consistent with the data on $A_N$ 
in $\Lambda$ production\cite{Bra98,E704b}. 
It explains in particular 
the surprising experimental result\cite{Bra98,E704b} that 
$\Lambda$ produced by a spin zero $(ud)$-diquark 
from the polarized projectile also 
exhibits left-right asymmetry\cite{BL96}! 

These two points are supplemented by 
the following two points, which are consistent with 
the recent ALEPH and OPAL data\cite{ALEPH,OPAL} on 
longitudinal polarization of $\Lambda$ 
in $e^+e^-\to Z^0\to q\bar q\to \Lambda+X$, 
in order to give a description of $P_H$ 
in hadron-hadron collisions.

(1) Quark polarization is {\it not} destroyed in 
fragmentation.

(2) The SU(6) wave-function can be used to describe 
the relation between the spin of the fragmenting quark and that of 
the hadron produced in the fragmentation process.

We recall that the SU(6) wave-functions 
have been widely used 
in studying hyperon polarization 
in hadron-hadron collisions in the literature, 
(see, e.g., [\ref{And79}], [\ref{DeG81}], and [\ref{Pon85}]),  
i.e. point (2) has been assumed to be true. 
From this point, we obtain immediately 
that $\Lambda$ polarization is completely 
determined by its $s$ valence quark.
This result is quite different from that suggested by 
the recent polarized deep-inelastic 
lepton-nucleon scattering (DIS) data \cite{Mall96}.  
The recent DIS data suggest that, at large $Q^2$, 
the quarks and antiquarks carry 
only a small fraction of the nucleon spin. 
When applied to $\Lambda$, 
it suggests that \cite{Jaffe96}
$\Lambda$ spin is not completely determined by its $s$ quark. 
This initiated the discussions 
(see, e.g. [\ref{Jaffe96}-\ref{Ellis96}]) 
whether such a picture of the  spin structure 
should also be used for describing the relation between 
the spins of the fragmenting quarks and 
the polarization of the hyperon produced 
in the fragmentation processes. 
It has been pointed out in [\ref{BL98}] 
that measurements 
of the longitudinal polarization of $\Lambda$ in 
$e^+e^-\to Z^0\to\Lambda X$ can provide some hints  
to answer this question. 
The available data\cite{ALEPH,OPAL}
are still far from accurate and enormous enough 
to provide a conclusive judgement. 
But these available data\cite{ALEPH,OPAL} 
seem to favor\cite{BL98} 
the SU(6) description.\cite{foot}

We note that the four points (a), (b), (1) and (2) 
form the basis of the picture in [\ref{LB97}].  
They are consistent with the data now 
available\cite{Bra98,E704a,E704b,ALEPH,OPAL}.  
Whether, if yes, how they can be derived 
from QCD are questions which have been discussed 
frequently in recent years in literature.  
Many models have been proposed   
to understand in particular point (a) and (b) in terms 
of quark-parton model in the framework of QCD.
Since the purpose of [\ref{LB97}] and that of the present paper 
are to discuss the close relationship between 
the left-right asymmetries observed in 
single-spin hadron-hadron collisions with 
hyperon polarization in 
unpolarized hadron-hadron collisions, we use 
these points  as input.  
In this sense, 
the results obtained in the following 
and those in [\ref{LB97}] are  consequences of 
a phenomenological model which 
is consistent with the basic principles and 
the empirical facts from other experiments.  
Using this picture, 
we showed \cite{LB97} that various data on hyperon 
polarization in inclusive production processes 
can be understood provided  
that the $s$ and $\bar s$ taking part 
in the associated production of 
the $\Lambda $ and $K^+$ 
have opposite transverse spins. 
We found in particular that, in this picture, 
$\Lambda$ polarization comes mainly from the 
$\Lambda$'s which contain 
two valence-quarks of the proton projectile 
and are associated with a spin zero meson such as $K^+$ 
which contains the remaining valence quark of the proton.

We stress that, although the model is 
consistent with that proposed by 
DeGrand and Miettinen \cite{DeG81} on some points, 
it differs very much from the latter. 
In [\ref{DeG81}], polarization of hyperon 
originates from a semi-classical effect, 
Thomas precession, which leads to the 
``slow-partons-spin-down-fast-partons-spin-up'' rule. 
This would in particular predict 
that hyperons produced in $e^+e^-$ annihilations should also 
be transversely polarized, which contradicts 
the data\cite{TASSO,ALEPH}. 
The model in [\ref{LB97}] relates 
hyperon polarization $P_H$ in unpolarized collisions 
with left-right asymmetries $A_N$ observed\cite{Bra98} 
in single-spin inclusive hadron production processes. 
Data on $A_N$ suggest a correlation 
between the polarization of the valence quark and 
the transverse moving direction of the produced hadron 
[Points (a) and (b) mentioned above].
The picture in [\ref{LB97}] shows that the same 
correlations (a) and (b) lead also to 
hyperon polarization in unpolarized collisions. 

We now come to the 
application of  the picture to the diffractive process 
$pp\to \Lambda K^+p$. 
We note that this is the simplest channel for 
the inclusive process $pp\to \Lambda X$ 
and it has the following peculiarities: 
First, unlike  in $pp\to \Lambda X$,  
$\Lambda$ has to contain 
two of the valence quarks from the colliding proton 
in this process.    
The associated spin zero $K^+$  contains the other valence quark. 
This means that, in this channel, we have only $\Lambda$ 
produced by mechanism (b)  
which, according to the picture in [\ref{LB97}], 
provides the largest polarization.
Second, there is no contribution from hyperon decay.
This means that there is no contamination from such 
decay processes to the $\Lambda$ polarization. 
In fact, this is the only channel 
where these conditions are completely fulfilled. 
We therefore expect that $\Lambda$ polarization should 
take its maximum in this process.  
This is consistent with the R608 observation\cite{R608} that 
$P_\Lambda$ in this process is much larger than that 
in $pp\to \Lambda X$. 

In order 
to estimate $P_\Lambda$ in this process quantitatively, 
we recall that $P_\Lambda $ is defined with respect to 
the production plane (see Fig. 1) and 
$P_\Lambda<0$ means that the 
$\Lambda$ has a large probability to be 
polarized in the $-\vec n$ direction, 
where $\vec n\propto \vec p_B\times \vec p_\Lambda$ 
is the normal of the production plan. 
According to point (b), 
if $\vec p_\Lambda$ is in the direction 
as shown in the figure (denoted by ``left"), 
$K^+$ should be going right, 
thus $\vec p_B\times \vec p_{K^+}$ should be in the 
opposite direction as $\vec n$. 
According to point (a), 
the $u$-valence quark should 
have a large probability to be polarized 
in $-\vec n$ (downwards) direction. 
The difference of the probability 
for this $u_v$ to be polarized in $-\vec n$ 
and that to be polarized in $\vec n$ direction 
is given by the constant $C$ mentioned in point (a). 
Since $K^+$ is a spin zero object, 
the $\bar s$ in $K^+$ should 
be polarized in $\vec n$ direction 
if $u_v$ is polarized in $-\vec n$ direction. 
Hence, the $s$ quark thus the $\Lambda$ should be polarized 
in $-\vec n$ direction if 
$s$ and $\bar s$ have opposite transverse spins. 
(See Fig.2).
We see, in this case, the polarization of $\Lambda$ 
is the same as that for the remaining $u$-valence quark
which is contained in the associatively produced $K^+$.
This implies that,
\begin{equation}
P_\Lambda (pp\to \Lambda K^+p)=-C.
\end{equation}
Using the value $C=0.6$ determined (see, e.g. [\ref{BL96}] 
and the references given there)
by fitting the $A_N$ data, 
we obtain that $P_\Lambda (pp\to \Lambda K^+p)=-0.6$ 
which is in good agreement with the R608 data\cite{R608} 
$P_\Lambda (pp\to \Lambda K^+p)=-0.62\pm0.04$. 

This result is rather encouraging. 
We therefore made a detailed analysis for the related 
spin effects in this processes.
We found that single-spin reaction 
$p(\uparrow)p\to \Lambda K^+p$ with 
transversely polarized proton $p(\uparrow)$
is an ideal place to test the picture proposed in [\ref{LB97}]
and its applicability to diffractive processes.  
We obtained  the following:

($\alpha$) There should be a 
large left-right asymmetry $A_N$ for 
$\Lambda$ as well as for  $K^+$ 
in $p(\uparrow)p\to \Lambda K^+p$ 
in the fragmentation region 
of the transversely polarized proton $p(\uparrow)$, and,
\begin{equation}
A_N^\Lambda [p(\uparrow)p\to \Lambda K^+p]=
-A_N^{K^+}[p(\uparrow)p\to \Lambda K^+p]=-C.
\end{equation}
This is because $|\Lambda^\uparrow>=(ud)_{0,0}s^\uparrow$, 
thus only the configuration $(ud)_{0,0}u^\uparrow$ in the projectile 
proton $p(\uparrow)$ contributes to 
the process $p(\uparrow)p\to \Lambda K^+p$. 
Hence, the $u$ valence quark contained 
in $K^+$ is upwards polarized.
According to the points (a) and (b) mentioned above, 
we obtain the results shown in Eq.(2).

($\beta$) The ``spin transfer parameter'' 
$D_{NN}^{\Lambda}$ for the produced $\Lambda$ 
should be positive and large 
in $p(\uparrow )p\to \Lambda K^+p$ 
in the fragmentation region 
of $p(\uparrow )$. 
We recall that $D_{NN}^{\Lambda}$ is defined as 
the probability for the produced $\Lambda$ 
to be polarized in the same transverse direction 
as the projectile. 
Although the $ud$ diquark which comes from the projectile 
to form the produced $\Lambda$ has to be in the spin-zero state 
thus carries no information of polarization of the projectile, 
the remaining $u$-valence quark  
completely determines the polarization of the projectile.  
They are polarized in the same direction. 
Hence, the $\bar s$ has to be polarized in the opposite direction 
as the projectile since $K^+$ has spin zero. 
The $s$ quark, which has opposite transverse spin as the $\bar s$, 
thus the $\Lambda$ containing this $s$-quark 
should therefore be polarized 
in the same transverse direction of the projectile. 
Hence, we obtain,
\begin{equation}
D_{NN}^{\Lambda}[p(\uparrow)p\to \Lambda K^+p]=1.
\end{equation}
Both ($\alpha$) and ($\beta$) are predictions 
which can be tested in future experiments.
The predictions 
for the process $pp\to \Lambda K^+p$ are summarized 
in Table I.

Here, it should be mentioned that 
$\Lambda$ polarization has recently been measured\cite{Felix96} 
in another exclusive process 
$pp\to p\Lambda K^+\pi^+\pi^-\pi^+\pi^-$
at incident momentum of 27.5 GeV/c. 
The results show  the following: 
(i) The magnitude of $P_\Lambda$ 
in this channel is much smaller than that in 
$pp\to p\Lambda K^+$.
(ii) $P_\Lambda$ is approximately 
the same for events where 
$K^+$ and $\Lambda$ are produced in the same hemisphere 
and for those where they are in the opposite hemispheres. 
We show that both (i) and (ii) are consistent  
with the picture mentioned above. 
First, unlike that in $pp\to p\Lambda K^+$, 
$\Lambda$ in this channel 
can be produced by two or one of 
the three valence quarks of the colliding proton. 
But, according to the picture \cite{LB97}, 
only the $\Lambda$'s 
produced by two valence quarks 
are polarized, those produced by one are not. 
Second, while $\Lambda$ in $pp\to \Lambda K^+p$ 
is definitely associated with $K^+$, 
in $pp\to p\Lambda K^+\pi^+\pi^-\pi^+\pi^-$, 
$\Lambda$ can be associated with 
vector meson such as $K^{*0}$ which 
subsequently decays into $K^+\pi^-$. 
As has been emphasized in [\ref{LB97}], 
the correlation between 
the $u$-valence quark in the associated 
meson and the $s$-quark in 
the produced $\Lambda$ will not be destroyed 
if more spin-zero mesons are associatively produced, 
(See Fig.3.), 
but it will be destroyed if (spin one) vector meson(s) 
is (are) involved. 
Hence, we expect that point (i) is true.
Furthermore, since $P_\Lambda$ is not changed if more 
spin zero mesons are produced in between (Fig.3), 
it is therefore unimportant whether 
$K^+$ and $\Lambda$ are produced 
in the same or in the opposite hemispheres. 
This implies (ii) should also be true.  

In summary, we have successfully applied the 
picture proposed in [\ref{LB97}] 
to $\Lambda$ production in diffractive processes. 
The obtained results 
are in agreement with the data now available.
Predictions have been given which can be tested 
in future experiments.
 

We are indebted to Professor Meng Ta-chung who initiated 
this research and participated in the early stage of this work. 
Part of the results in this paper are taken from an 
unpublished note (Ref.[\ref{BLM}]) 
written together with him.
We thank R. Rittel, Wang Qun and Xie Qu-bing 
for helpful discussions.
This work was supported in part by 
the National Natural Science Foundation of China (NSFC), 
the State Education Commission of China, and 
the Australian Research Council.

\newpage

\begin{table}
\caption{Predictions on different spin parameters in 
the diffractive process $pp\to \Lambda K^+p$ 
at high energies of the picture in [\ref{LB97}]. 
The data for $P_\Lambda[pp\to\Lambda K^+p]$ 
is taken from [\ref{R608}].}
\begin{tabular}{lll}
\hline
 & theory & data \\
$P_\Lambda[pp\to\Lambda K^+p]$ & $-C(=-0.6)$ & $-0.62\pm0.04$\\
$A_N^\Lambda[p(\uparrow)p\to\Lambda K^+p]$ & $-C(=-0.6)$ \\
$A_N^{K^+}[p(\uparrow)p\to\Lambda K^+p]$ & $C(=0.6)$ \\
$D_{NN}^\Lambda[p(\uparrow)p\to\Lambda K^+p]$ & $\ \ \ \ \ 1$ &\\
\hline
\end{tabular}
\end{table}
\nopagebreak

\begin{figure}[t]
\psfig{file=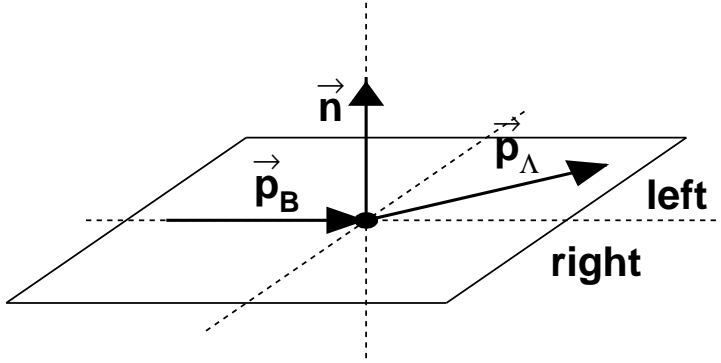,height=10cm}
\caption{Hyperon polarization is defined 
with respect to the  unit vector 
$\vec n\equiv \vec p_B\times \vec p_\Lambda/|\vec p_B\times \vec p_\Lambda|$ 
which is perpendicular to  the production plane.  
Here, $\vec p_B$ and $\vec p_\Lambda$ are momentum of the beam hadron 
and that of the produced $\Lambda$, respectively.  
From the figure, we see in particular that $\vec n$ is pointing upwards if 
$\vec p_\Lambda$ is pointing to the left.}
\end{figure} 

\begin{figure}[t]
\psfig{file=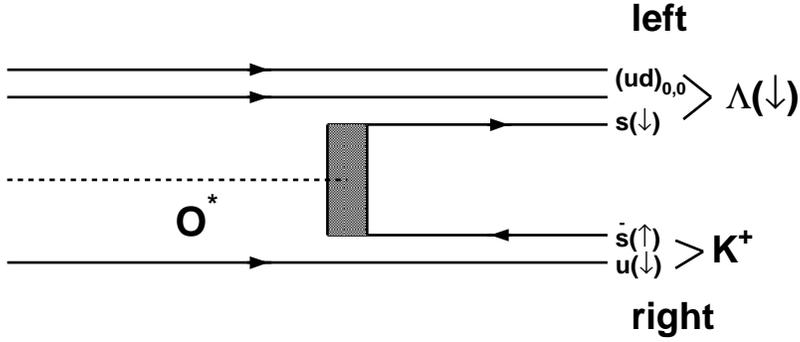,height=10cm}
\caption{Associated production of $\Lambda$ containing two of 
the valence quarks of the projectile proton and $K^+$ containing 
the remaining $u$ valence quark. 
Here, the three long straight solid lines represent 
the three valence quarks from the projectile proton; 
$O^*$ represents some unknown object which carries the 
energy-momentum needed to create the $s\bar s$-pair. 
From the figure, we see 
that polarization of $\Lambda$  
and that of the remaining $u$ valence quark are 
the same provided that the $s$ and $\bar s$ have opposite spins.}
\end{figure} 

\begin{figure}[t]
\psfig{file=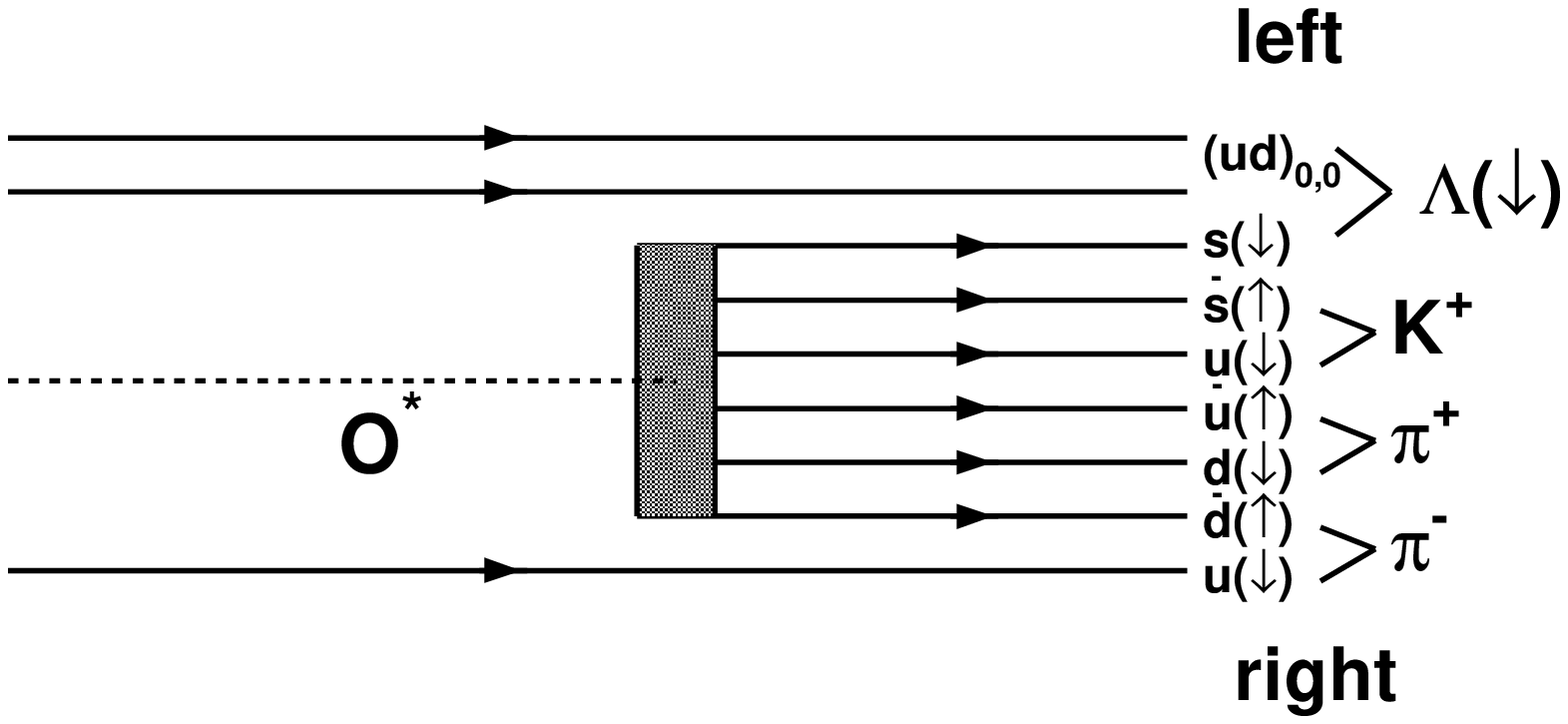,height=10cm}
\caption{Production $\Lambda$ containing two of 
the valence quarks of the projectile proton associated with 
more than one pseudoscalar mesons with one of them containing 
the remaining $u$ valence quark. 
We see here  that the correlation between 
the polarization of $\Lambda$  
and that of the remaining $u$ valence quark is 
not destroyed if more spin zero mesons are 
associatively produced. (C.f. Fig.2).}
\end{figure} 


\begin {thebibliography}{99}
\bibitem{foot} We note that this does not necessary mean that 
only constituent quarks are involved in the collisions.
We recall that the SU(6) baryon wavefunction 
is a result of the following requirements:
A baryon is a composite of three 
spin $1/2$ objects (the quarks), and 
these three quarks form a color singlet 
(thus have a completely 
antisymmetric color wave function) 
so that their flavor and spin wave functions 
have to be completely symmetric. 
Hence, the validity of such wave function in 
describing the relationship between 
the spins of the fragmenting quarks 
and that of the produced baryon may imply that 
these quarks first evolve into constituent quarks 
then combine into the baryon. 
It can also imply that  
they first combine to form  
the $|qqq\rangle$ fock state of the baryon 
then evolve into the complete physical baryon, and so on.
Both possibilities are consistent 
with the presently popular fragmentation models.
\end{thebibliography}
\end{document}